\documentclass{elsarticlemod}

\usepackage[english]{babel}
\usepackage{hyperref}
\usepackage{graphicx}
\usepackage{amsbsy}

\usepackage{subfigure}
\usepackage{amsmath,amsfonts}

\newif\ifdraftversion
\draftversionfalse

\newif\ifusetikz
\usetikzfalse

\ifusetikz

  \usepackage{tikz}
  \usepackage{pgfplots}
  \usepackage{shellesc}
  \usetikzlibrary{decorations.pathreplacing,external}
  \usepgfplotslibrary{statistics}
  \colorlet{bigsmallcolor}{green}
  \colorlet{monodispersecolor}{red}
  \colorlet{decreasingcolor}{orange}
  \colorlet{increasingcolor}{violet}
  \colorlet{bellcolor}{blue}
  \colorlet{modalfivecolor}{brown}
  \colorlet{smallbigcolor}{cyan}
  \colorlet{uniformcolor}{black!30!green}
  \colorlet{darkgreen}{black!50!green}
  \colorlet{modaltwocolor}{magenta}

  \pgfplotsset{standard/.style={only marks,thin,black,mark=x,mark options={scale=0.5},forget plot}}
  \pgfplotsset{bigsmall/.style={only marks,mark=*,bigsmallcolor,mark options={scale=0.75}}}
  \pgfplotsset{monodisperse/.style={only marks,mark=square*,monodispersecolor,mark options={scale=0.75}}}
  \pgfplotsset{decreasing/.style={only marks,mark=triangle*,decreasingcolor,every mark/.append style={rotate={270}}}}
  \pgfplotsset{increasing/.style={only marks,mark=triangle*,increasingcolor,every mark/.append style={rotate={180}}}}
  \pgfplotsset{bell/.style={only marks,mark=triangle*,bellcolor,every mark/.append style={rotate={90}}}}
  \pgfplotsset{modal5/.style={only marks,mark=star,modalfivecolor,every mark/.append style={rotate={45}}}}
  \pgfplotsset{smallbig/.style={only marks,mark=diamond*,smallbigcolor}}
  \pgfplotsset{uniform/.style={only marks,mark=triangle*,uniformcolor}}
  \pgfplotsset{modal2/.style={only marks,mark=pentagon*,modaltwocolor,mark options={scale=0.75}}}
  
  \pgfplotsset{bigsmallvoid/.style={only marks,mark=o,bigsmallcolor,mark options={scale=0.75}}}
  \pgfplotsset{monodispersevoid/.style={only marks,mark=square,monodispersecolor,mark options={scale=0.75}}}
  \pgfplotsset{decreasingvoid/.style={only marks,mark=triangle,decreasingcolor,every mark/.append style={rotate={270}}}}
  \pgfplotsset{increasingvoid/.style={only marks,mark=triangle,increasingcolor,every mark/.append style={rotate={180}}}}
  \pgfplotsset{bellvoid/.style={only marks,mark=triangle,bellcolor,every mark/.append style={rotate={90}}}}
  \pgfplotsset{modal5void/.style={only marks,mark=star,modalfivecolor,every mark/.append style={rotate={45}},mark options={scale=0.75}}}
  \pgfplotsset{smallbigvoid/.style={only marks,mark=diamond,smallbigcolor}}
  \pgfplotsset{uniformvoid/.style={only marks,mark=triangle,uniformcolor}}
  \pgfplotsset{modal2void/.style={only marks,mark=pentagon,modaltwocolor,mark options={scale=0.75}}}

  \newcommand\includeexternaltikzfigure[3]{\tikzsetnextfilename{#3/#2}\input{#1/#2}}
\else
  \newcommand\includeexternaltikzfigure[3]{\includegraphics{#3/#2}}
\fi

\ifdraftversion
  \usepackage[colorinlistoftodos, obeyFinal]{todonotes}
  \newcommand\pablo[1]{\ifusetikz\tikzexternaldisable\fi \todo[inline,color=red!50!white]{Pablo commented: #1} \ifusetikz\tikzexternalenable\fi}
  \newcommand\nacho[1]{\ifusetikz\tikzexternaldisable\fi \todo[inline,color=blue!50!white]{Nacho commented: #1} \ifusetikz\tikzexternalenable\fi}
\else
  \newcommand\pablo[1]{}
  \newcommand\nacho[1]{}
\fi

\begin{document}

\begin{frontmatter}

\title{Use of Machine Learning for unraveling hidden correlations between Particle Size Distributions and the Mechanical Behavior of Granular Materials}


\author[UPMaddress]{Ignacio G. Tejada\corref{mycorrespondingauthor}}\cortext[mycorrespondingauthor]{Corresponding author}
\ead{ignacio.gtejada@upm.es}

\author[EPFLaddress]{Pablo Antolin}
\ead{pablo.antolin@epfl.ch}

\address[UPMaddress]{Universidad Polit\'ecnica de Madrid, ETSI de Caminos, Canales y Puertos\\ C/ Profesor Aranguren 3, 28040 Madrid, Spain}
\address[EPFLaddress]{\'Ecole Polytechnique F\'ed\'erale de Lausanne, Institute of Mathematics\\ CH-1015 Lausanne, Switzerland}


\begin{abstract}
A data-driven framework was used to predict the macroscopic mechanical behavior of dense packings of polydisperse granular materials.
The Discrete Element Method, DEM, was used to generate $92,378$ sphere packings that covered many different kinds of particle size distributions, PSD, lying within 2 particle sizes.
These packings were subjected to triaxial compression and the corresponding stress-strain curves were fitted to Duncan-Chang hyperbolic models. 
A multivariate statistical analysis  was unsuccessful to relate the model parameters with common geotechnical and statistical descriptors derived from the PSD.
In contrast, an artificial Neural Network (NN) scheme, trained with a few hundred DEM simulations, was able to anticipate the value of the model parameters for all these PSDs, with considerable accuracy.
This was achieved in spite of the presence of noise in the training data.
The NN revealed the existence of hidden correlations between PSD of granular materials and their macroscopic mechanical behavior.
\end{abstract}


\end{frontmatter}


\section{Introduction}

The specific values of properties such as strength, compressibility and permeability of dry and cohesionless coarse grain materials (including sand, gravel, railway ballast or rockfill) depend on the features of the constituent particles (\textit{intrinsic properties}) and on the way in which the particles are arranged (\textit{state parameters}).
Among the intrinsic properties of a sand, the surface friction, the compressibility and the strength of individual grains,  the particle shape and particle size distributions are known to play a crucial role in its macroscopic properties~\cite{Sullivan2002,Santamarina2004,Cavarretta10,Guida2018}.
Relative density and confining pressure are the most influent state variables for dry granular soils~\cite{Been1991} and govern the mechanical behavior of the material to a large extent~\cite{Rowe1962,Thornton2000,Andersen2013}. 

The relationship between the particle size distribution, PSD, and the mechanical behavior is not yet fully understood.
On one hand, the effects of variations in the PSD are not independent from those produced by variations of other intrinsic properties or state parameters.
For example, the state parameter $\psi$, proposed within the theoretical framework of the critical state of sands~\cite{Been1991}, helps to distinguish between the contractive or dilatant behavior exhibited by a sand upon triaxial compression. 
However the critical state line, and hence the value of $\psi$ associated to given void ratio $e$, changes with the PSD~\cite{Jiang2018}.
As another example, there is a complex interplay between size and shape polydispersity, as shown by numerical modeling~\cite{Nguyen2015}.
On the other hand, linking single quantities (ma\-xi\-mum and minimum dry density, critical state void ratio, macroscopic friction angle, stiffness, etc.) to a PSD  is not immediate, since the latter is a highly variable curve that is many times long-tailed and/or multi-modal.
Descriptors derived from the PSD are not enough to anticipate macroscopic (void ratio, stiffness, friction angle) or microscopic features (average coordination number, fraction of non-contributing particles, etc.) obtained after a given process. Neither geotechnical descriptors, such as the $D_{xx}$ (\textit{i.e.}, the sieve size passed by $xx$ percent in weight of the sample), the coefficient of curvature $C_\text{c}$ or the uniformity coefficient $C_\text{u}$, nor statistical descriptors (mean, variance, skewness, kurtosis, etc.) enable satisfying estimations.
There is not clear procedure to work directly with the whole PSD curve.
Even in the case of very idealized systems (\textit{e.g.},~packings of spheres) variations of the PSD may lead to considerably differences in the fabric resulting after a packing protocol~\cite{Shaebani12,Wiacek2014}, in the  relative density~\cite{Guida2018} or in the shear strength~\cite{Goncu2013}.
In the case of non-idealized systems this can be even worse, as several kinds of physico-chemical phenomena occur on different length and time scales.
Relationships between geotechnical descriptors obtained from the PSD and geotechnical properties have been sought (\textit{e.g.},~\cite{Wichtmann2013,Monkul2016,Xiao2016}), but findings are always empirical and limited to a specific set of soils and stress paths. 

The use of large datasets enables promising techniques to understand how the complex behavior of granular systems can be anticipated from the microscopic features.
For example, the use machine learning techniques, together with complex network theory, has allowed for the establishment of relationships between the fabric of a packing and some macroscopic geotechnical properties, such as the  permeability~\cite{vanDerLinden2016,Kamrava2020} or the effective heat transfer coefficient~\cite{Fei2020}.
These techniques were also applied in \cite{lai2019reconstructing} for obtaining morphological information of granular materials, including their particle size distribution, from X-ray computed tomography images.
The use of artificial neural networks, or just Neural Networks (NN), has been proposed as a potentially useful technique to model materials behavior~\cite{Ghaboussi1991,Pernot1991}.
In the case of  geotechnical applications, NNs have been used for unsaturated soils (to predict the shear strength~\cite{Lee2003}, to model their mechanical behavior~\cite{Johari2011} and to determine the effective stress parameter~\cite{Ajdari2012}), for fine-grain soils (to predict the compression index from granulometry and index properties~\cite{Park2011}), for rocks (to predict the uniaxial compressive strength and the elastic modulus~\cite{Dehghan2010}) and for coarse-grain soils --sands and gravels-- to model the mechanical behavior ~\cite{Ellis1995,Penumadu1999,Banimahd2005}.
The  inputs for these NN approaches included both intrinsic properties and state parameters.
In some of these cases the target outputs were directly some model parameters (namely, the compression index~\cite{Park2011}, the apparent cohesion~\cite{Lee2003}, the effective stress parameter~\cite{Ajdari2012} or the elastic modulus and unidimensional compression strength~\cite{Dehghan2010}). 
For all the other cases above mentioned (\textit{i.e.},~\cite{Ellis1995,Penumadu1999,Banimahd2005}), the purpose of NNs was to reproduce the stress-strain curve by anticipating new values of stress or strains obtained when some others were changed in a controlled way.
The datasets were the result of a limited number of laboratory experiments (around several tens to a few hundred).
Only in~\cite{Park2011}, the database included data from near 1 thousand experiments. In some cases, a single stress-strain curve measured in a laboratory experiment was used to gather the data.
NNs have also been been successfully used to link parameters used in DEM simulation to the macroscopic behavior of granular materials (angle of repose~\cite{Benvenuti2016} or hopper discharge rate~\cite{Kumar2018}). 
These researches have shown how NNs enable for the understanding of the bulk behavior of granular materials with a reduced number of virtual experiments.
 
In this research the role played by the PSD in the mechanical behavior of an idealized system of polydisperse spheres has been investigated by means of massive numerical testing with the DEM and NNs.
This approach may shed light on the mechanical behavior of dry coarse-grain soils.
This is a very timely moment since new techniques, such as X-ray tomography, are allowing for an exhaustive characterization of the microstructure of granular packings (including PSD and fabric)~\cite{wiebicke2020measuring}.
There are two considerable differences with respect to the previously referenced works.
On one hand, the NN is built on a dataset that was the outcome of a series of more than $90$ thousand virtual experiments, performed with samples of varying PSD.
The set of PSDs is the outcome of a systematic exploration of possible cases lying within two particle sizes.
The probability and size increments used during a discretization of the sample space determined the number of PSDs to explore.
We simulated all the cases to have a sufficiently large data sample, to find out how the accuracy of the estimations depends on the size of the training dataset and to know what the Probability Distribution Functions, PDF, of the target outputs for the NN are.
On the other hand, we focus on the role played by the PSD in the macroscopic behavior.
We used quite simple granular systems (made of elastic and frictional spheres) and experiments to exclusively focus on the differences in the mechanical behavior due to the PSD.

The proposed approach is \textit{ab initio} as phenomenological laws are not used (except that for the contact mechanics interaction).
Neither intrinsic parameters that cannot be defined on the grain scale (such as maximum or minimum dry density, etc.) nor state parameters related to packing features (void ratio, average coordination number, etc.) were introduced.
The mechanical features of particles and the packing and compression protocols have always been the same and the only difference from one case to another is the PSD.
Albeit the simplicity of the systems, non-linear and stress dependent stress-strain curves were observed (showing the typical behavior of loose sands), with non-obvious variations from one case to another.
The data were fitted to the celebrated Duncan-Chang hyperbolic model~\cite{Duncan1970}, which is defined by two model parameters, namely, the tangent elastic modulus $E_\text{0}$ and the ultimate deviatoric stress $\sigma_\text{ult}$.

Thus, the proposed NN receives as input a discrete description of the PSD of a granular material at hand, and returns as output $E_\text{0}$ and $\sigma_\text{ult}$.
As it will be illustrated below, the network is able to predict the Duncan-Chang model's parameters with a high accuracy, extremely fast, and even in the presence of noisy training data.
Indeed, it proved itself to be a powerful tool for unraveling the existing correlations between PSD of granular materials and their macroscopic mechanical behavior, hidden to the naked eye.

The rest of this paper is structured as follows: Initially, the discrete element method used for the generation of virtual triaxial experiments, as well as the considered PSDs and the obtained results are described in Section \ref{sec:dem}; secondly, in Section \ref{sec:ann}, we present the basic principles of artificial neural networks, together with the design of the networks used in this work and their training process; the results obtained with the NNs are presented and discussed in Section \ref{sec:results}, as well as a study of the amount of required data to train them and their robustness with respect to noisy data; finally, conclusions are drawn in Section \ref{sec:conclusions}.

\section{Massive DEM triaxial testing} \label{sec:dem}

The discrete element method~\cite{Cundall1979}, DEM, has been proven to be a very effective tool for the study of the macroscopic mechanical behavior of  granular materials under drained  \cite{Thornton2000,ng2004triaxial,Sibille2007,kozicki2009numerical,salot2009influence,kozicki2014discrete,Zhou2017,Xie2017,Jiang2020,Xu2020} and undrained~\cite{Gong12,Jiang2018b,Salimi2020} triaxial or biaxial compression.
In this work the DEM is used to perform virtual drained triaxial tests for a large number of sphere packings with different PSDs.
In what follows we describe the model used for carrying out such simulations, as well as the obtained results.

\subsection{Numerical setup} \label{sec:numerical_setup}
We performed $92,378$  DEM simulations of triaxial compression tests on samples made of particles following varying PSDs.
The different PSDs used in each case were selected according to a systematic exploration described as follows: Particle diameters ranged between $D_{\text{min}}=0.05$ m and $D_{\text{max}}=0.15$ m; this interval was divided into $10$ equal size bins $\left(D_i, D_i + \Delta D\right]$ with $\Delta D = \left( D_{\text{max}} - D_{\text{min}}\right) / 10$, $D_0 = D_{\text{min}}$ and $i=0, 1, \dots, 9$.
 The central size of each bin is $d_i = D_i + 0.5 \Delta D$.
The expected percentage in mass of the particles within each size bin $i$ is denoted as $p_i$.
We consider that $p_i$ is a discrete variable that can take values from $0.0$ to $1.0$ and spaced by $0.1$.
All possible combinations $\left\{ p_i \right\}^9_{i=0}$ satisfying $\sum_{i=0}^{9} p_i = 1.0$ are considered.
This procedure led to the $92,378$ cases of PSDs that were subsequently used in the triaxial tests.

\begin{figure}
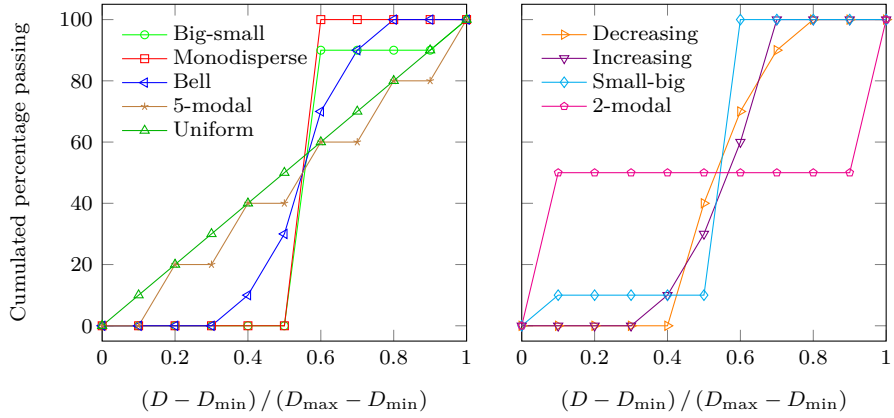
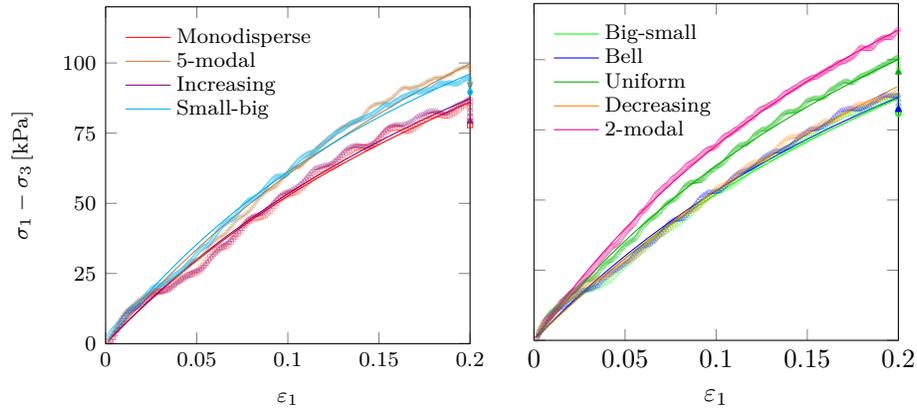

	\centering
	\subfigure[Special particle size distributions]
  {\includeexternaltikzfigure{figures}{PSDs1}{.}\hspace{3mm}\includeexternaltikzfigure{figures}{PSDs2}{.}\label{fig:PSDs}}\\
	\subfigure[Strain-stress virtual curves (markers) and Duncan-Chang model adjustment (continuous line)]
  {\includeexternaltikzfigure{figures}{stressstrain1}{.}\hspace{3mm}\includeexternaltikzfigure{figures}{stressstrain2}{.}\label{fig:stressstrain}}\\
	\caption{$9$ special PSDs (out of $92,378$) were selected for illustrative purposes.
	The upper figures show the cumulated percentage passings.
	The figures below show the stress-strain curves obtained through virtual testing}
	\label{fig:special_cases}
\end{figure}

Once all the PSDs were defined, a random sample of particles was generated for each of them.
The mass of the particles was uniformly distributed in each bin.
The considered set of PSDs includes very different kinds of granular systems: Monodisperse, well graded, gap-graded multimodal distributions, etc.
A few of them, which could be more recognizable by readers, have been particularly considered for illustrative purposes.
These special PSDs are labeled and shown in Fig.~\ref{fig:PSDs}.

\begin{figure}
	\centering
	\subfigure[Initial state]{\includegraphics[width=0.32\textwidth]{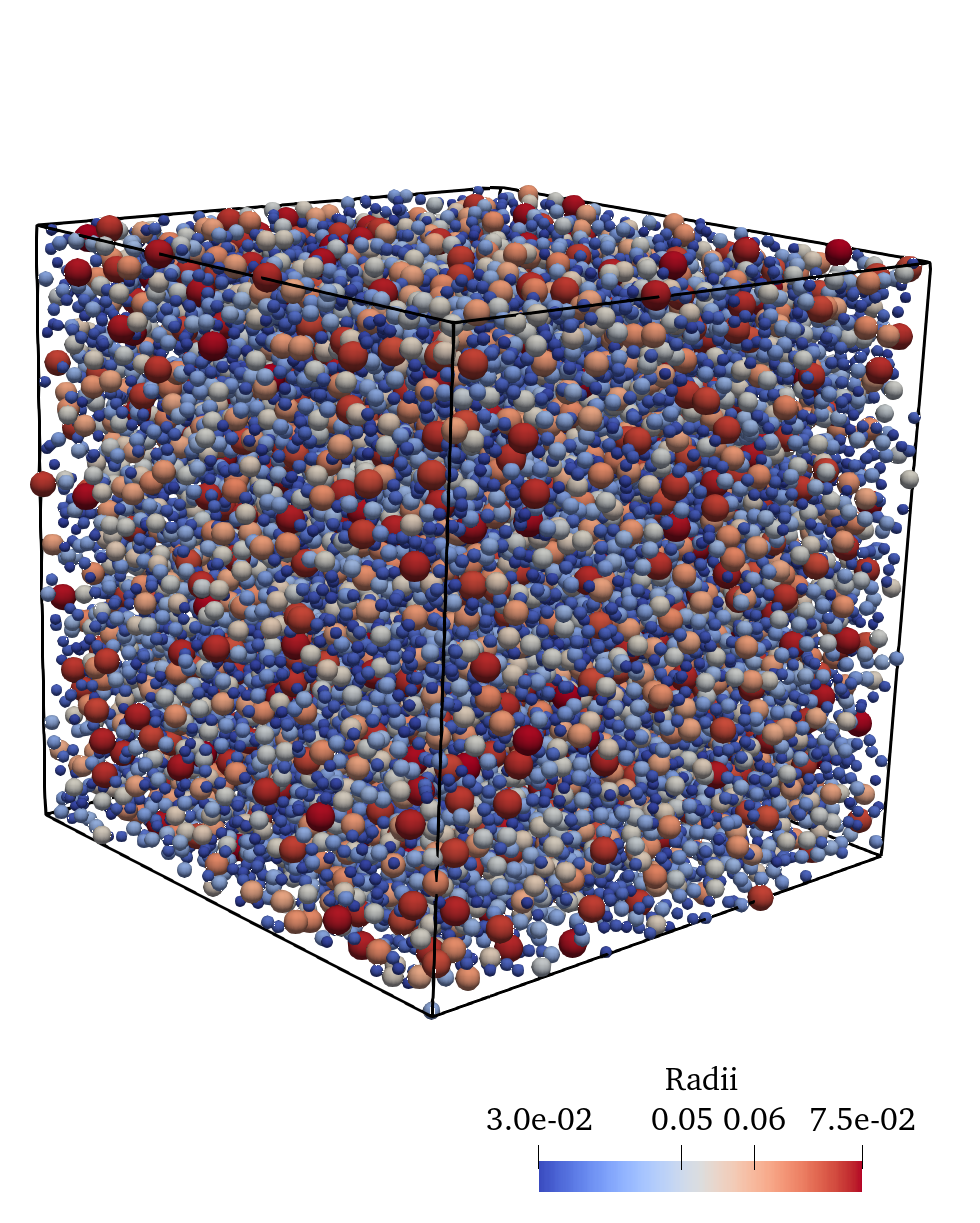}\label{fig:dem3Da}}\hfill
	\subfigure[Isotropic compression state]{\includegraphics[width=0.32\textwidth]{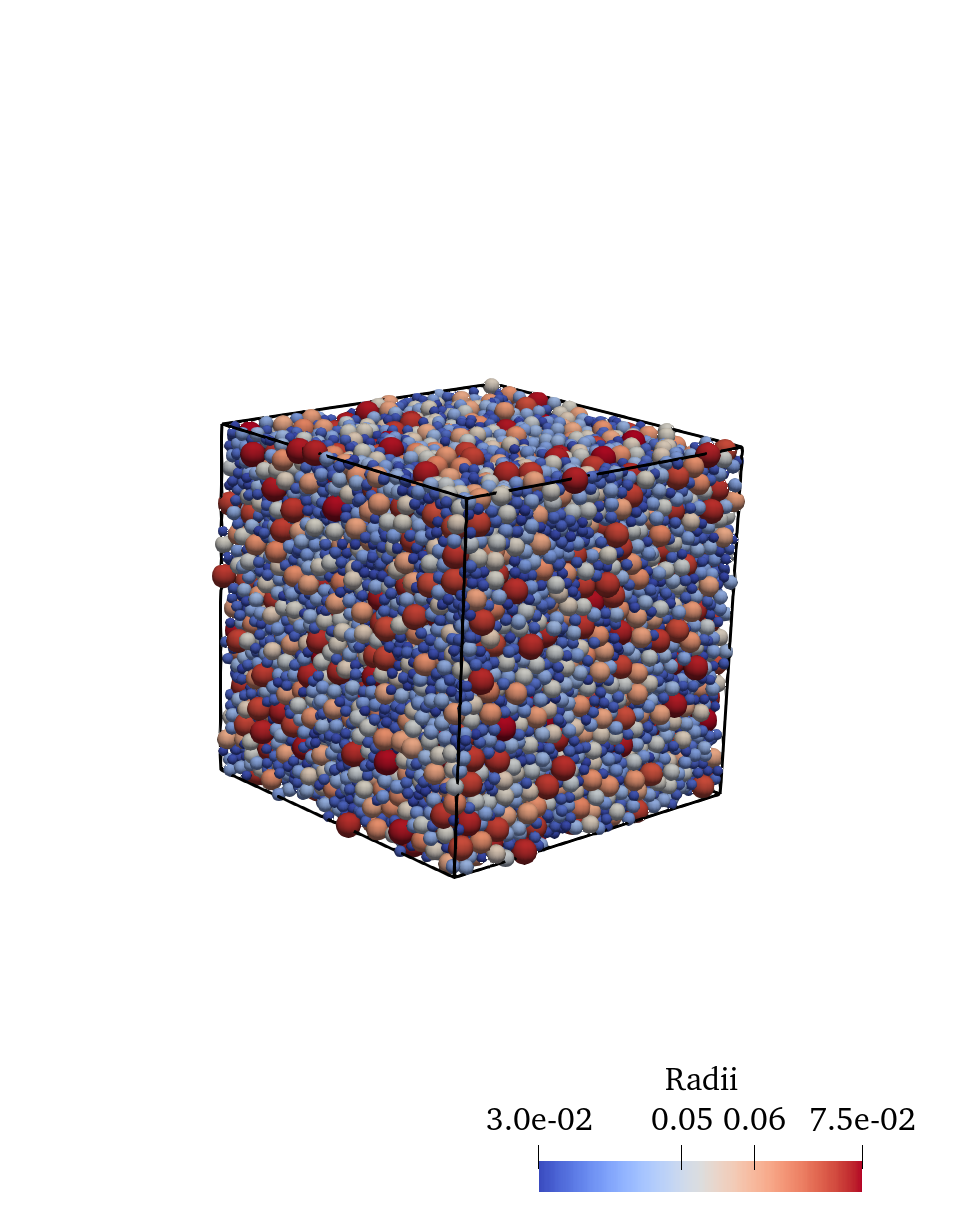}\label{fig:dem3Db}}\hfill
	\subfigure[Final state]{\includegraphics[width=0.32\textwidth]{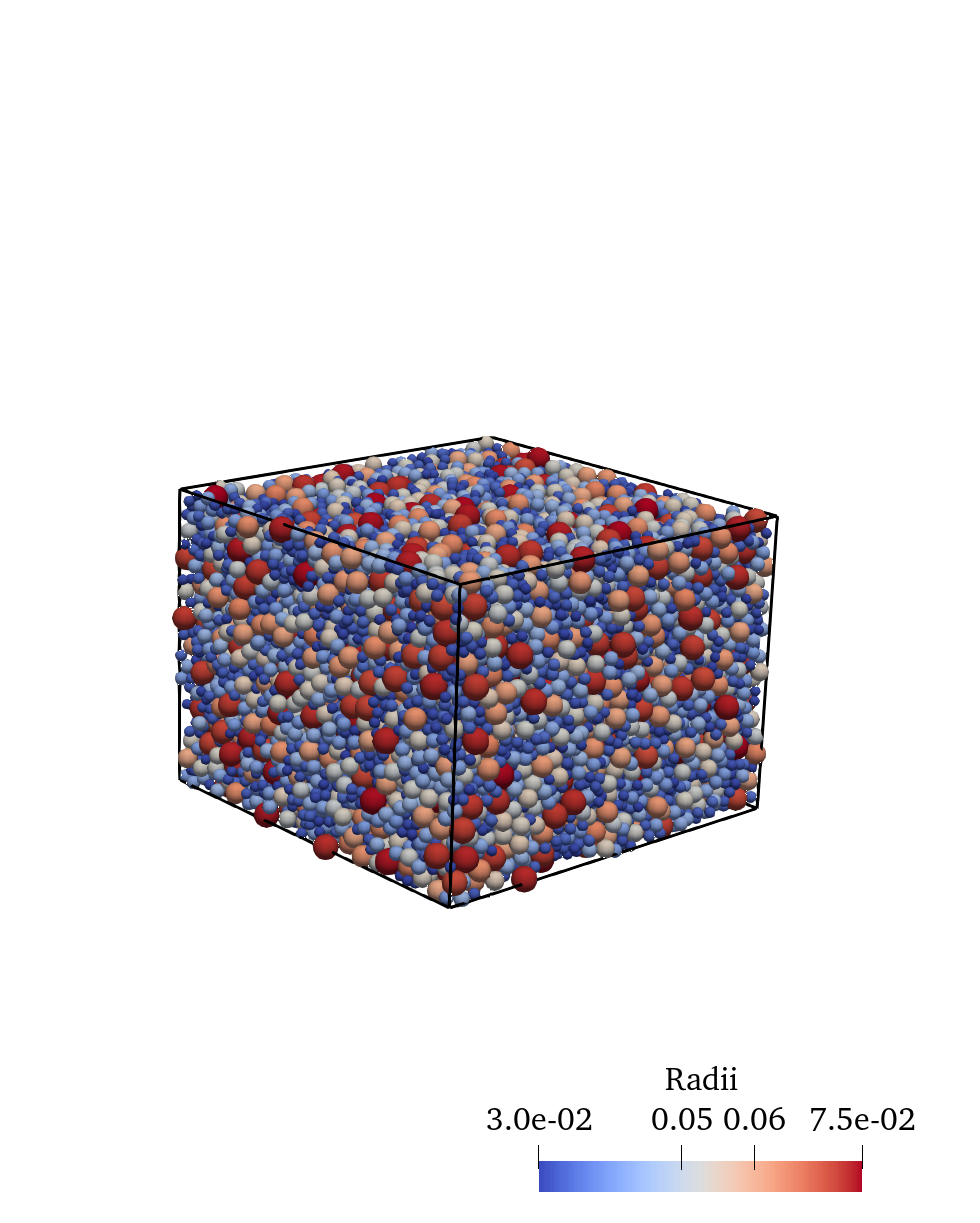}\label{fig:dem3Dc}}\\
	\caption{3D Models of YADE-DEM illustrating the steps of numerical experiments: 1) A random loose cloud of around $20,000$ particles is located within a box; 2) the simulation box is reduced to achieve a packing that is in equilibrium under isotropic stress; and 3) the simulation box is reduced in one direction while the stress is maintained in 2 perpendicular directions.}
	\label{fig:dem3D}
\end{figure}

For each PSD, a sample was generated by randomly locating a loose cloud of around $20,000$ spherical particles within a cubic box (Fig.~\ref{fig:dem3Da}).
We imposed periodic boundary conditions and then the cubic box was isotropically shrunk to achieve a dense packing under isotropic compression conditions $\sigma_1 = \sigma_2= \sigma_3 = 100$ kPa, where $\sigma_1$, $\sigma_2$ and $\sigma_3$ are the principal stresses (Fig.~\ref{fig:dem3Db}).
Then the stress was kept in 2 perpendicular directions ($\sigma_2$ and $\sigma_3$), while the sample was shortened in the third perpendicular direction until reaching a unit strain $\varepsilon_1 = 0.2$ (Fig.~\ref{fig:dem3Dc}). 
The corresponding average stress $\sigma_1$ was measured at several strain levels.
The deviatoric stress-strain curve, $\sigma_d=\sigma_1-\sigma_3$ vs $\varepsilon_1$, was registered and fitted to a Duncan-Chang hyperbolic model~\cite{Duncan1970}, which is defined by 2 model parameters, namely, the tangent elastic modulus, $E_\text{0}$ and the ultimate deviatoric stress $\left( \sigma_1 - \sigma_3 \right)_\text{ult} = \sigma_\text{ult}$:
\begin{align}
\label{eq:DuncanChang}
\sigma_d = \left( \sigma_1 - \sigma_3 \right) = \frac{\varepsilon_1}{\frac{1}{E_\text{0}}+\frac{\varepsilon_1}{\sigma_\text{ult}}} \,\text{.}
\end{align}
A few examples of generated curves (corresponding to the special PSDs) can be seen in Fig.~\ref{fig:stressstrain}.

\subsection{Numerical model}
We used the DEM implemented in YADE-DEM~\cite{yade}~\footnote{www.yade-dem.org}.
Particles behave as rigid solids that obey the laws of classical mechanics.
The interaction between particles is produced through a soft contact model.
In particular, we used a simple linear elastic and frictional contact law.
This is a common choice in DEM simulation~\cite{Cundall1979,Herrmann1998}.
Normal forces between particles are thus computed as
\begin{align}
\mathbf{F}_{\text{n},ij} = k_\text{n} \delta_{ij} \mathbf{n}_{ij} \, \text{,}
\end{align}
where $\mathbf{F}_{\text{n}, ij}$ is the normal force exerted by particle $j$ on particle $i$, $\delta_{ij} = r_{ij}  - \left( R_i + R_j \right)$ is the distance overlap, $R_i$ and $R_j$ are the particles' radii, $\mathbf{r}_{ij}$ is their relative position vector, $\mathbf{n}_{ij} = \mathbf{r}_{ij} /  \left\lVert\mathbf{r}_{ij}\right\rVert $ is its associated unit vector, and $k_\text{n}$ is the normal contact stiffness.
In this model, $k_\text{n}$ was related to the Young's modulus of the material, $E = 1.0$ GPa, as $k_\text{n} =  2 E R_i R_j / \left( R_i + R_j \right)$.

If two particles that were previously in contact (\textit{i.e.}, $\delta_{ij}<0$) are displaced in a direction $\boldsymbol{\xi}_{ij} /\xi_{ij}$ perpendicular to $\mathbf{n}_{ij}$, an opposite shear force appears.
Shear forces are limited by the inter-particle friction:
\begin{align}
\mathbf{F}_{\text{s},ij} =  - \min{\left( k_\text{s} \xi_{ij}  ,\, \tan{\phi} F_{\text{n},ij} \right)}\, \frac{\boldsymbol{\xi}_{ij}}{\xi_{ij}}  \,\text{,}
\end{align}
where $\mathbf{F}_{\text{s}, ij}$ is the shear force exerted by particle $j$ on particle $i$, $\boldsymbol{\xi}_{ij}$ is the total tangential displacement of the contact, $\phi = \Pi/6$ radians is the inter-particle friction angle and $k_\text{s} = 0.25 k_\text{n}$ is the shear stiffness.

The density of particles $\rho = 10^6$ kg/m$^3$ (as the size of the particles and the stiffness) was scaled to reduce the collision time and therefore the critical timestep used in the explicit integration of the equations of motion.
The maximum strain rate imposed during the triaxial compression was fixed according to this critical timestep and updated on the fly to speedup simulations.
A numerical damping was used to dissipate the kinetic energy.
Details can be found in~\cite{yade}.
\begin{figure}[h!]
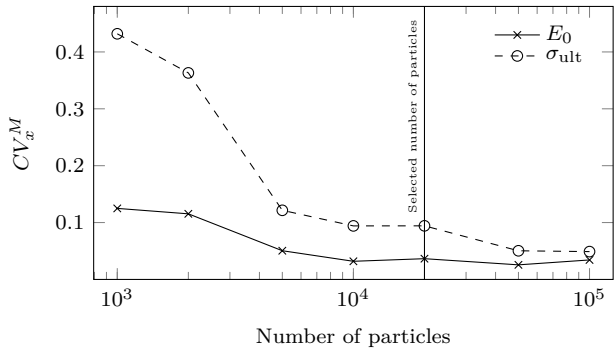

	\centering
	\includeexternaltikzfigure{figures}{CoVvsN}{.}
	\caption{Coefficient of variation of the parameters for the Duncan-Chang hyperbolic model as a function of the number of particles in the sample.
	Samples followed the uniform PSD in Fig.\ref{fig:PSDs}.
	The experiment was repeated 15 times for each number of particles}
	\label{fig:CoV}       
\end{figure}
\subsection{Precision and performance} \label{sec:precision}
As the generated samples include a finite number of particles, the computed stress-strain curves for a single PSD may fluctuate around the expected values.
Accordingly, the values of the Duncang-Chang model parameters  obtained from a single DEM triaxial test, are only punctual estimations,  $E_{0,\left[\text{DEM}\right]}$, $ \sigma_\text{ult,[\text{DEM}]}$, which are generally different from the expected values, $\bar{E}_\text{0}$ and $\bar{\sigma}_\text{ult}$.
There are several reasons for this variability: The size of the particles used in each simulation is randomly chosen according to the PSD, particles are randomly located within the simulation box and the system is chaotic.
In any case, the larger the sample, the smaller the fluctuation. 
The expected variability of measurements was assessed through a series of virtual triaxial tests.
These tests were performed with samples made of varying number of particles but that always followed the same PSD (the uniform PSD in Fig.~\ref{fig:special_cases}).
The experiment was repeated $15$ times for each number of particles to gather a statistical sample of $E_0$ and $\sigma_{\text{ult}}$ values.
A coefficient of variation  was defined for each model parameter $x$ as $CV^M_x = s_x / \bar{x}$ (where $s_x$ is the sample standard deviation, $\bar{x}$ is the sample mean and $M$ stands for measurement).
Results are shown in Fig.~\ref{fig:CoV}.
In order to achieve a good compromise between accuracy and computational cost, the size of samples was limited to around $20,000$ particles in the DEM experiments used to train the NN.
With this number of particles the $CV^M$ of $E_\text{0}$ and $\sigma_\text{ult}$ are expected to remain around $0.05$ and $0.10$, respectively (see Fig.~\ref{fig:CoV}).  
These numbers indicate that the measurement of $E_\text{0}$ is more precise than that of $\sigma_\text{ult}$.

The numerical experiments were computed using the version 2018.02b of YADE-DEM~\cite{yade}, running on Ubuntu 18.04.4 64 bits, on a server machine with four processors Intel Xeon Gold 6148 $2.40$ GHz, with 20 physical cores each, and 1 TB of RAM memory.
As a rough estimation, each single DEM simulation took on average 1 hour and 20 minutes on a single core.
Therefore, the total computation time for processing the $92,378$ samples was around $5,135$ days.
In order to speed up the process, many computer cores were used for running multiple independent simulations in parallel.
Thus, the total process time was reduced to 4 and a half months of computation, approximately.

\subsection{Virtual triaxial testing results} \label{sec:dem_results}
\begin{figure}[h!b]
	\centering
	\subfigure[Histogram of $E_\text{0}$]
	{\includeexternaltikzfigure{figures}{histogram_E0}{.}\label{fig:E0histogram}}\hfill
	\subfigure[Histogram of $\sigma_\text{ult}$]
	{\includeexternaltikzfigure{figures}{histogram_Qult}{.}\label{fig:q0histogram}}\\
	\subfigure[$E_\text{0}$ vs.\ $\sigma_\text{ult}$]
	{\includegraphics{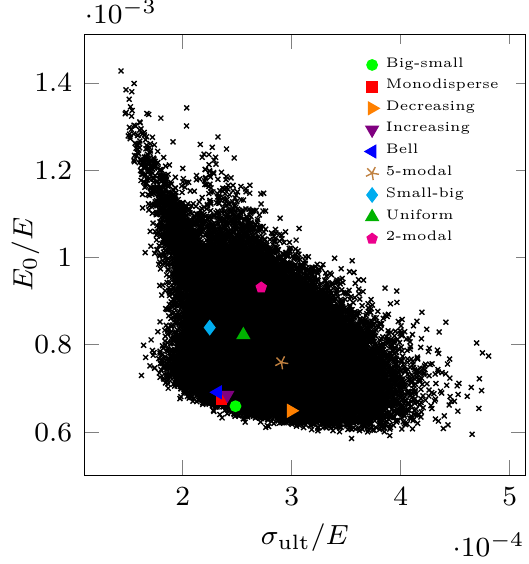}\label{fig:EvsQult}}
	\caption{Histograms of $E_\text{0}$ and $\sigma_\text{ult}$ values obtained from virtual triaxial testing with the set of $92,378$ PSDs explored,
	and variation between both values.
	The cases reported in Fig.~\ref{fig:special_cases} are highlighted, while their actual values are gathered in Table~\ref{tab:values}}
	\label{fig:Evs}
\end{figure}

The $92,378$ samples were virtually subjected to triaxial compression.
The corresponding stress-strain curves presented the typical behavior of loose sands.
A good matching between each series of data and a Duncan-Chang hyperbolic curve was achieved.
The values of $E_0$ obtained from DEM after a flat sampling over the set of PSDs, are distributed as shown in Fig.~\ref{fig:E0histogram}.
The sample mean is $\bar{E}_\text{0} / E = 7.83 \cdot 10^{-4}$, its standard deviation is $s_{E_\text{0}} / E = 8.59 \cdot 10^{-5}$ and the maximum and minimum values are $E_{\text{0},\text{max}} / \bar{E}_\text{0} = 1.82$ and $E_{\text{0},\text{min}} / \bar{E}_\text{0} = 0.75$, respectively.
The coefficient of variation of this problem quantifies how the expected value of a specific PSD may separate from the mean value across all the PSDs.
Regarding the tangent elastic modulus, the coefficient of variation is $CV_{E_\text{0}} = 0.11$.
With respect to the values of $\sigma_\text{ult}$ obtained from DEM, the distribution is shown in Fig.~\ref{fig:q0histogram}, the sample mean is $\bar{\sigma}_\text{ult} / E = 2.81 \cdot 10^{-4}$, its standard deviation is $s_{\sigma_\text{ult}} / E = 3.72 \cdot 10^{-5}$ ($CV_{\sigma_\text{ult}} = 0.13$) and the maximum and minimum values are $\sigma_{\text{ult},\text{max}} / \bar{\sigma}_\text{ult} = 1.71$ and $\sigma_{\text{ult},\text{min}} / \bar{\sigma}_\text{ult} = 0.51$, respectively.

\begin{figure}[h!b]
	\centering
	\subfigure[$E_\text{0}$ vs.\ particles' mean diameter]
	{\includegraphics{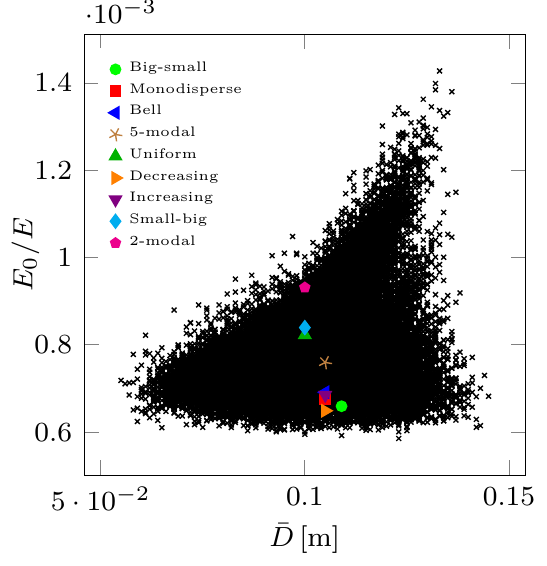}\label{fig:EvsMean}}
	\subfigure[$E_\text{0}$ vs.\ standard deviation]
	{\includegraphics{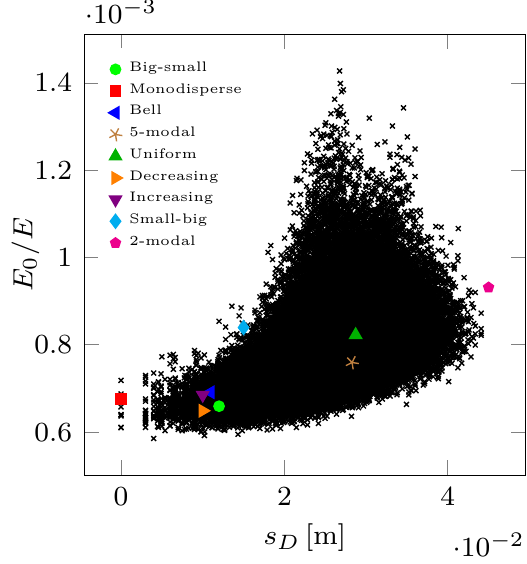}\label{fig:EvsSTD}}
	\subfigure[$E_\text{0}$ vs.\ skewness]
	{\includegraphics{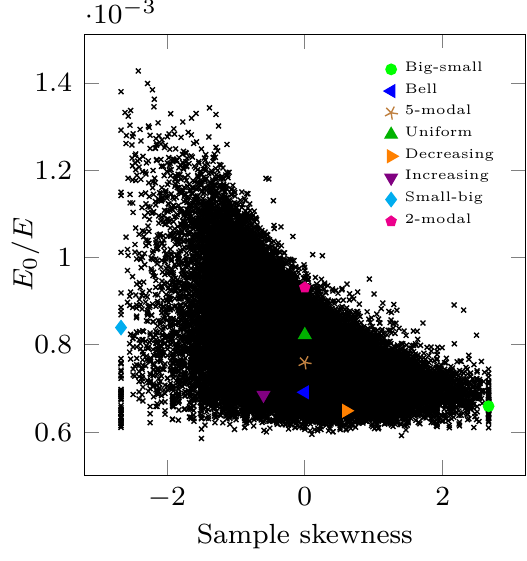}\label{fig:EvsSkewness}}
	\subfigure[$E_\text{0}$ vs.\ excess kurtosis]
	{\includegraphics{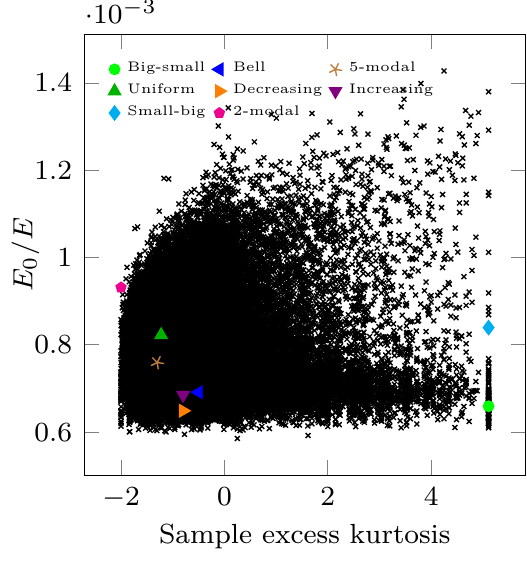}\label{fig:EvsKurtosis}}
  \caption{Variation of $E_\text{0}$, obtained from virtual triaxial testing, compared to different statistical descriptors, namely the mean diameter of particles, the standard deviation, skewness and excess kurtosis of the PSD (see Table~\ref{tab:descriptors}).
	The cases reported in Fig.~\ref{fig:special_cases} are highlighted, while their actual values are gathered in Table~\ref{tab:values}}
  \label{fig:EvsStatistical}
\end{figure}

\begin{figure}[h!t]
	\centering
  \subfigure[$E_\text{0}$ vs.\ curvature]
	{\includegraphics{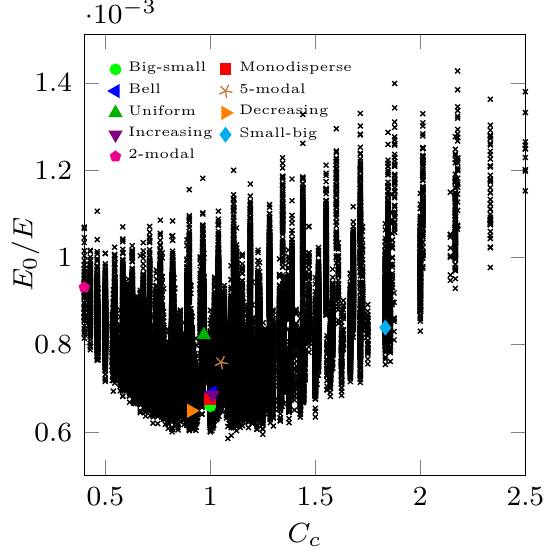}\label{fig:EvsCc}}
  \subfigure[$E_\text{0}$ vs.\ uniformity]
	{\includegraphics{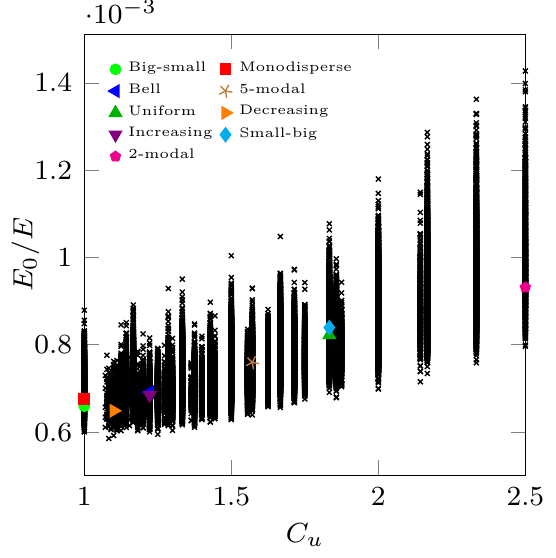}\label{fig:EvsCu}}
  \caption{Variation of $E_\text{0}$, obtained with virtual triaxial testing, with geotechnical descriptors of the PSD, namely the curvature and the uniformity (see Table~\ref{tab:descriptors}).
	The cases reported in Fig.~\ref{fig:special_cases} are highlighted, while their actual values are gathered in Table~\ref{tab:values}}
  \label{fig:EvsGeotechnical}
\end{figure}

These results evidence that variations of the Duncan-Chang model parameters can be found depending on the PSD.
Unfortunately, there is no sign of correlation between the two model parameters (see Fig.~\ref{fig:EvsQult}).
In addition, they neither correlate to the set of inspected statistical or geotechnical descriptors described in Table~\ref{tab:descriptors}, as it can be observed in Figs.~\ref{fig:EvsStatistical} and \ref{fig:EvsGeotechnical}, respectively. 

\begin{table}[h!b]
	\small
	\centering
	\begin{tabular}{lcl}
		\hline
		 Descriptor & Symbol & Definition\\
		\hline
         &\textit{Geotechnical descriptors} &\\
		Uniformity coefficient & $C_\text{u}$ & $C_\text{u} = \frac{D_{60}}{D_{10}}$\\
		Coefficient of curvature & $C_\text{c}$ & $C_\text{c} = \frac{D_{30}^2}{ D_{10} D_{60}}$\\
	 &\textit{Statistical descriptors} &\\
		Expected value  & $\bar{D}$ & $\bar{D} = \sum_i p_i d_i$ \\
		Standard deviation  & $s_D$ & $s_D = \sqrt{\sum_i p_i \left(d_i - \bar{D} \right)^2}$ \\
		Skewness  & $\tilde{\mu}_3$ & $\tilde{\mu}_3  = \frac{\sum_i p_i \left(d_i - \bar{D} \right)^3}{s_D^3}$ \\
		Excess Kurtosis  & $\text{K}_{\left[ D \right]} - 3$ & $\text{K}_{\left[ D \right]}-3= \frac{\sum_i p_i \left(d_i - \bar{D} \right)^4}{s_D^4} - 3$ \\
		\hline
	\end{tabular}
	\caption{Set of descriptors used to relate the parameters of the Duncan-Chang model to the PSD. $d_i = D_i + 0.5 \Delta D$}
	\label{tab:descriptors}
\end{table}
For the 9 special PSDs included in Fig.~\ref{fig:special_cases}, the values of the considered statistical and geotechnical descriptors are gathered in Table~\ref{tab:values} and also shown in Figs.~\ref{fig:EvsStatistical} and \ref{fig:EvsGeotechnical}.
%

\begin{table}[h!b]
\small
\centering
\begin{tabular}{lccrrrrrr}
\hline
PSD & \multicolumn{1}{c}{$\hat E_\text{0}$} & \multicolumn{1}{c}{$\hat \sigma_\text{ult}$} & \multicolumn{1}{c}{$\bar{D}$} & \multicolumn{1}{c}{$s_D$} & \multicolumn{1}{c}{Skew.} & \multicolumn{1}{c}{Kurt.} & \multicolumn{1}{c}{$C_\text{u}$} & \multicolumn{1}{c}{$C_\text{c}$} \\
\hline
\small{Big-small}  & $6.59$ & $2.48$ & $0.109$ & $0.01$ & $2.66$  & $5.11$  &  $1.0$ & $1.00$ \\
\small{Monodisperse}    & $6.77$ & $2.35$ & $0.105$ & $0.00$ & ---     & ---     &  $1.0$ & $1.00$ \\
\small{Decreasing} & $6.49$ & $2.99$ & $0.105$ & $0.01$ & $0.60$  & $-0.80$ &  $1.1$ & $0.91$ \\
\small{Increasing} & $6.85$ & $2.41$ & $0.105$ & $0.01$ & $-0.60$ & $-0.80$ &  $1.2$ & $1.01$ \\
\small{Bell}     & $6.91$ & $2.32$ & $0.105$ & $0.01$ & $0.00$  & $-0.50$ &  $1.2$ & $1.01$ \\
\small{5-modal}    & $7.60$ & $2.91$ & $0.105$ & $0.03$ & $0.00$  & $-1.30$ &  $1.6$ & $1.05$ \\
\small{Small-big}  & $8.39$ & $2.25$ & $0.100$ & $0.02$ & $-2.66$ & $5.11$  &  $1.8$ & $1.83$ \\
\small{Uniform} & $8.22$ & $2.56$ & $0.100$ & $0.05$ & $0.00$  & $-2.00$ &  $1.8$ & $0.97$ \\
\small{2-modal}    & $9.31$ & $2.72$ & $0.100$ & $0.05$ & $0.00$  & $-2.00$ & $2.5$ & $0.40$ \\
\hline
\end{tabular}
\caption{Values of $E_\text{0}$, $\sigma_\text{ult}$, obtained with virtual triaxial simulations, and other indicators for the cases in Fig.~\ref{fig:special_cases}.
The descriptors included in the table, correspond to the mean $\bar{D}$, standard deviation $s_D$, skewness and excess kurtosis of the particle diameters; and the curvature $C_\text{c}$ and uniformity $C_\text{u}$ coefficients (geotechnical indicators). $\hat E_\text{0} = E_\text{0}/E\text{$\times10^4$}$ and $\hat \sigma_\text{ult} = \sigma_\text{ult}/E\text{$\times10^4$}$. These values are presented graphically in Figs.~\ref{fig:Evs}, \ref{fig:EvsStatistical} and \ref{fig:EvsGeotechnical}}
\label{tab:values}
\end{table}
In the light of these results, to establish relationships between PSD descriptors and Duncan Chang model parameters does not seem feasible.

\section{Artificial neural networks} \label{sec:ann}

Artificial neural networks, or simply Neural Networks (NN), are biologically inspired computing systems able to learn from data.
Data abundance, together with increasing computing power, are probably the two main factors behind the great success of these algorithms and their exponential growth during the last decade, despite the fact that their origin dates back to the early 40s of 20th century \cite{haykin1994neural}.
Artificial neural networks, together with other machine learning techniques, have been proven very successful tools for tackling tasks as image recognition, language processing or financial forecasting, to name just a few.
Beyond doubt, machine learning in general, and neural networks in particular, are powerful tools for untangling complex patterns on large datasets.

Motivated by the apparent lack of correlation between PSD descriptors and the Duncan Chang model parameters evidenced in the previous section, 
in this work we present, as an accurate alternative, the use of NNs for inferring the macroscopic mechanical behavior of polydisperse granular packings.
As it will be seen in the results presented in Section \ref{sec:results}, this tool will help us to find hidden connections between the particle size distribution of spherical packings and their macroscopic mechanical behavior.

\subsection{The multilayer perceptron}

\begin{figure}[ht]
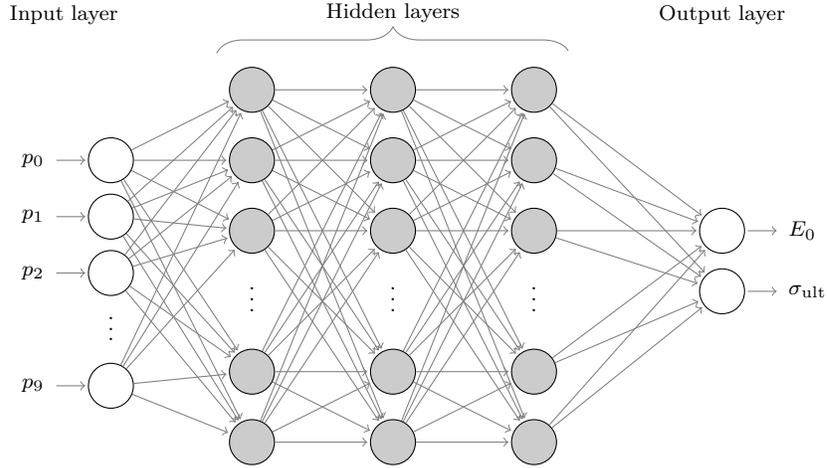

\centering
\includeexternaltikzfigure{figures}{nn_sketch}{.}
\caption{Multilayer perceptron architecture}
\label{fig:mlp}
\end{figure}

One of the most simple and commonly used NN architectures is the Multi-Layer Perceptron (MLP).
The MLP can be seen as a  non-linear function that maps input data to output data.
It consists of several layers: One input layer, one or more (intermediate) hidden layers, and one output layer.
The input information is feed-forwarded from the input layer, through all the intermediate layers,  up to the output layer.
Each layer is composed of one or more nodes (or neurons) that are the basic computational units (see Fig.~\ref{fig:mlp}).
At each layer, the neurons are fed with the output generated by the neurons of the previous layer, they process the data, filter it through a non-linear activation function, and produce new output values that feed the neurons of the next layer (if any).
The presence of non-linear activation functions grants NNs the ability of approximating non-trivial functions.
Indeed, as stated by the universal approximation theorem \cite{cybenko1989}, feed-forward NNs with a single (finite) hidden layer and differentiable activation functions, can approximate any continuous function; and in the case of two hidden layers or more, any function \cite{cybenko1988}.

Let us describe how a MLP generates output values from given input.
Let $L+1$ be the number of layers in a NN, such that $L \in\mathbb{Z}^+$ and $L > 1$, and let $N^{(l)}\in\mathbb{Z}^+$ be the number or neurons of the $l$-th layer, with $l=0,\dots, L$,
where the layer $0$ corresponds to the input layer and the $L$-th layer is the output one.
We denote as $x^{(l)}\in\mathbb{R}^{N^{(l)}}$ the input vector of the $l$-th layer and, accordingly, $x^{(l+1)}\in\mathbb{R}^{N^{(l+1)}}$ is the output of that layer and the input of the next one.
Thus, $x^{(0)}$ are the input values of the network and $x^{(L)}$ are the output ones.
Starting from input vector $x^{(0)}$, the values of layers $1$ to $L$ are computed through the recursive expression:
\begin{align} \label{eq:mlp}
x^{(l+1)} = \varphi\left(W^{(l)}\,x^{(l)} + b^{(l)} \right) \,\text{,}
\end{align}
where $W^{(l)}\in\mathbb{R}^{N^{(l+1)}\times N^{(l)}}$ and $b^{(l)}\in\mathbb{R}^{N^{(l+1)}}$ are the weights matrix and bias vector, $W^{(l)}\,x^{(l)}$ is a matrix-vector product that results in a vector of length $N^{(l+1)}$ and  $\varphi:\mathbb{R}\to\mathbb{R}$ is the non-linear activation function.
Among others, the Rectified Linear Unit (ReLU) function \cite{nair2010rectified}, defined as $\varphi(z)=\max(0,z)$, is one of the most commonly used activation functions. 
In the case in which $z$ is a vector, as it is the case of Eq.~\eqref{eq:mlp}, $\varphi$ is applied to each vector component independently.

On the other hand, the coefficients of the weights matrix $W^{(l)}$ and the bias vector $b^{(l)}$ are a collection of trainable parameters that describe the NN.
Those parameters, initially unknown, are determined by means of a process known as training.
The goal of the training is to find a set of values for those parameters that leads to an accurate input-output mapping of the network for the training dataset (a subset of the available input-output samples).
Finding the locally optimal parameters is a minimization process of a (loss) function that measures the distance (in a certain norm) between the known output sample values and the ones predicted by the network.
The mean squared error norm, used in this work, is one of the most commonly used loss functions.
Such optimization is commonly carried out by means of gradient-based iterative algorithms, like the ones of the family of stochastic gradient descendent methods, as it is the case of Adam \cite{kingma2014adam}.

For an in-depth discussion of MLPs and other NN architectures we refer the interested reader to \cite{haykin1994neural,Goodfellow-et-al-2016}.

\subsection{NNs for predicting Duncan-Chang model's parameters from PSDs} \label{sec:nn_architecture}

In this work we used MLP networks for predicting the parameters of the Duncan-Chang's model from a discrete description of the particle size distribution of a given spherical packing.
The use of MLP networks in the context of DEM simulation has allowed for the establishment of links between model features and bulk material behavior~\cite{Benvenuti2016}.
The definition, training and evaluation of the NNs was implemented using TensorFlow \cite{tensorflow2015-whitepaper}.
Thus, as it can be seen in Fig.~\ref{fig:mlp}, the designed network receives as input the ten PSD related values $\left\{ p_i \right\}^9_{i=0}$, defined in Section \ref{sec:numerical_setup}, and returns as output $E_\text{0}$ and $\sigma_\text{ult}$. Therefore, the input and output layers present 10 and 2 neurons, respectively.

The best network architecture (number of hidden layers and neurons) for the problem at hand is unknown \emph{a priori}.
Thus, in order to choose a good architecture, we systematically explored network configurations with different number of hidden layers and neurons per layer.
In the results presented in Section \ref{sec:results}, networks with 1, 2, 3 and 4 hidden layers and 8, 16, 32 or 64 neurons each (16 different architectures) were considered.
For all of them, the ReLU activation function was used for all the layers, including the output one.

The available virtual triaxials dataset ($92,378$ samples), was divided into three separated groups: the test dataset, composed of $72\%$ of the total samples available ($66,152$); the cross-validation dataset, $8\%$ of the total samples ($7,390$); and the test dataset, constituted by the remaining $20\%$ samples ($18,476$).
These three datasets were chosen randomly, nevertheless, they remain constant along the different analyses performed.
Whereas the test dataset was used for training the NNs, the cross-validation dataset helped us to compare the networks' performance and verifying the absence of undesired overfitting effects during the training process.
Finally, the test dataset was used for measuring the accuracy of the chosen networks when predicting a series of cases that were not used during the training process.

The training process of all the networks was carried out using Adam \cite{kingma2014adam} with 1000 epochs (training iterations through the whole test dataset).
And, in order to speedup the training process, the input and output data were previously normalized. Three different learning rates were considered for Adam, namely $\alpha=\left\{10^{-2},\, 10^{-3},\, 10^{-4}\right\}$.
The network's training is an inherently random process for two main reasons: Adam is by definition a stochastic algorithm in which the training samples are processed in a random order at each iteration; and the network's weights are randomly initialized.
Thus, in order to overcome these sources of randomness, each one of the 16 network architectures was trained 5 times for each learning rate.

After this training process, the network with the lowest loss function value for the cross-validation dataset was chosen.
No large differences were observed among the different architectures, nevertheless, a NN with a single hidden layer and 32 neurons on that layer presented a slightly better performance (network NN1 in Table~\ref{tab:nns}).

\begin{table}
\centering
\begin{tabular}{cccc}
\hline
Name & \# hidden layers & \# neurons per h.\ layer & $r$ \\
\hline
NN1 & 1 & 32 & 100\\
NN2 & 4 & 8 & 10\\
NN3 & 4 & 8 & 5\\
NN4 & 1 & 8 & 1\\
NN5 & 1 & 8 & 0.5\\
NN6 & 1 & 8 & 0.2\\
NN7 & 1 & 8 & 0.1\\
\hline
\end{tabular}
\caption{Different neural networks architectures considered in this work. Each NN consists of a different number of (\#) hidden layers and neurons per hidden layer, while the input and output layers have 10 and 2 neurons, respectively. Each network in the table was trained with a different number of samples ($r$ denotes the \% of the full test dataset)}
\label{tab:nns}
\end{table}

\section{Prediction of Duncan-Chang model parameters through neural networks} \label{sec:results}

The ability of the NNs described in Section \ref{sec:nn_architecture} to predict the values of $E_\text{0}$ and $\sigma_\text{ult}$ from PSD information, is discussed in this section.
In order to assess the prediction ability we consider discrepancies between NN predictions and DEM measurements for the same PSD.
The relative discrepancy for the model parameter $x$ ($E_\text{0}$ or $\sigma_\text{ult}$) associated to a PSD is evaluated according to:
\begin{align}
\label{eq:discrepancy}
\Delta_x = \frac{x_\text{[DEM]} -  x_\text{[NN]}}{x_\text{[DEM]}} \,\text{,}
\end{align}
where $x_\text{[DEM]}$ is the DEM measurement and $x_\text{[NN]}$ is the NN estimation.\\
In contrast to discrepancies, errors are defined with respect to the expected value of each model parameter, $\bar{x}$ ($\bar{E}_\text{0}$ or $\bar{\sigma}_\text{ult}$), associated to a PSD. However the expected values are usually unknown. They would be obtained with an infinitely large sample or by averaging over many random realizations of the triaxial test with packings following the same PSD.

As mentioned above, a randomly chosen test dataset of $20\%$ of the sample cases ($18,476$ out of $92,378$) was used for testing the network's accuracy.
These data are new to the network, in the sense that they were used neither during the training nor the cross-validation processes.
As presented below, different networks were trained using varying number of samples, nevertheless, the test dataset used for evaluating the networks' performance was kept constant along all the cases considered.

\subsection{Neural network ability to predict the Duncan-Chang model parameters} \label{sec:noise}
\begin{figure}[t!bh]
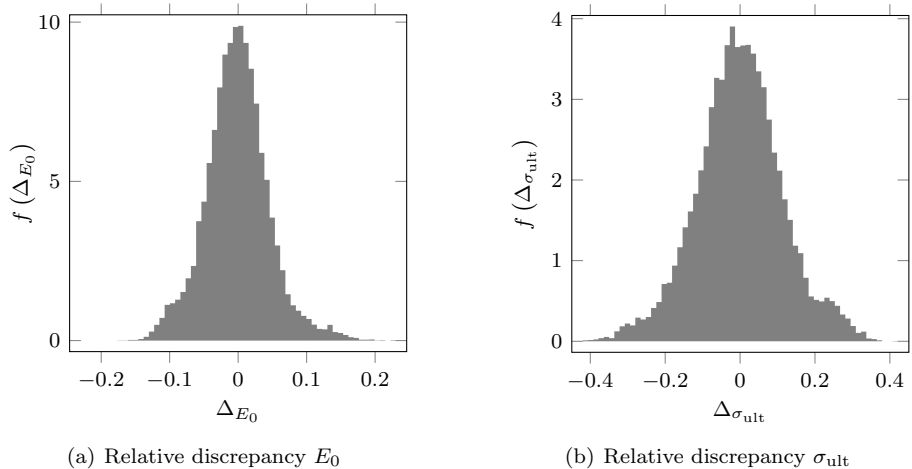

	\centering
	\subfigure[Relative discrepancy $E_0$]{\includeexternaltikzfigure{figures}{histogram_error_E0}{.}}\hfill
	\subfigure[Relative discrepancy $\sigma_\text{ult}$]{\includeexternaltikzfigure{figures}{histogram_error_Qult}{.}}
	\caption{Histogram of the relative discrepancies between NN and DEM in the estimation of Duncan-Chang model parameters when 80\% of the experiments were used to train the network. Results obtained with the network NN1 (see Table~\ref{tab:nns})}
	\label{fig:Discrepancies}
\end{figure}

Let us consider the network NN1 (see Table~\ref{tab:nns}), trained using the full test dataset (the 80\% of the $92,378$ cases, see Section \ref{sec:nn_architecture}).
For this specific NN the corresponding distributions of relative discrepancies within the test dataset are shown in Fig.~\ref{fig:Discrepancies}.
The NN1 anticipated values with discrepancies that fluctuated around 0.
The standard deviations were  $s_{\Delta_{E_\text{0}}} = 0.046$ and $s_{\Delta_{\sigma_\text{ult}}} = 0.115$.
The accuracy with the tangent elastic modulus is higher than with the ultimate deviatoric stress. 
This is consistent with the fact that the precision in the measurement is higher and the observed variation accross the considered cases is lower in the case of $E_\text{0}$.

For the sake of comparison, if the outcomes of the NN had been the sample means or random values, then the average discrepancies would have been considerably higher.
We checked this by using random estimations that followed either the observed probability distribution functions, PDFs, of $E_\text{0}$ and $\sigma_\text{ult}$ (Figs.~\ref{fig:E0histogram} and~\ref{fig:q0histogram}), or that followed uniform distributions lying between $E_\text{0,min}$ and $E_\text{0,max}$ (or between $\sigma_\text{ult,min}$ and  $\sigma_\text{ult,max}$).
This is summarized in Table~\ref{tab:NNvsRND}.
These results evidence the ability of the NN to predict the model parameters.
\begin{table}
	\centering
	\begin{tabular}{ccccc}
		\hline
		 & \multicolumn{4}{c}{Estimation}\\
		 & NN1 & Expected values & Observed PDFs & Uniform PDFs\\
		\hline
		$s_{\Delta_{E_\text{0}}}$ & $0.046$ & $0.106$ & $0.157$ & $0.342$\\
		$s_{\Delta_{\sigma_\text{ult}}}$ & $0.115$ & $0.156$ & $0.220$& $0.396$\\
		\hline
	\end{tabular}
	\caption{Comparison of the standard deviation of discrepancies with DEM for NN1 predictions and some random estimations}
	\label{tab:NNvsRND}
\end{table}

To correctly assess the accuracy of the NN, it is worth recalling that the data are pretty noisy (DEM measurements with a coefficient of variation for measurements of $CV^M_{E_\text{0}} \simeq 0.05$ and $CV^M_{\sigma_\text{ult}} \simeq 0.10$, as seen in Section \ref{sec:precision}).
It is also important to mention that the PDFs of $E_\text{0}$ and $\sigma_\text{ult}$, when all PSDs are considered, are bell-shaped (\emph{cf.\ }Figs.~\ref{fig:E0histogram} and~\ref{fig:q0histogram}) with $CV_{E_\text{0}} \simeq 0.11$ and $CV_{\sigma_\text{ult}} \simeq 0.13$, respectively.
Thus, despite the narrow margin left by the measurement precision and the distribution of values associated to this problem, the NN anticipated the Duncang-Chang model parameters from the PSD with the same accuracy than the precision of the DEM experiments with which it was trained.

This fact unveils the existence of hidden correlations between the PSD and the macroscopic mechanical behavior of granular materials, that are encoded in the NN, and, in the light of the results presented Section~\ref{sec:dem_results}, seemed hidden.
This result is even more interesting taking into account the fact that the DEM data used for training the NN are noisy, as discussed below in Section \ref{sec:noise}.

\subsection{Neural network accuracy with respect to the size of the DEM training dataset} \label{sec:noise}
Once we knew the expected accuracy of the NN predictions, we progressively reduced the size of the training datasets.
Along all the presented results, the test cases were always the same 20\% subset of the total.
Our goal was to estimate the number of DEM tests (out of the $73,902$ possible) that are needed to effectively train the NN without significantly compromising its accuracy.
To assess the accuracy of a NN trained with a certain subset of the training dataset, we evaluated the network for the test dataset and computed the mean squared deviations, $MSD_{r \left[ x \right]}$, where $x$ refers to the model parameter (either $E_\text{0}$ or $\sigma_\text{ult}$).
The subscript $r$ denotes the percentage of the potential training cases (out of the $73,902$ possible) that were used in each training set.
\emph{E.g.}, $r=10\%$ means that only $7,390$ samples of the test dataset  were used to train the network.
$MSD_{r \left[ x \right]}$ is defined as:
\begin{align}
\label{eq:MSD}
MSD_{r \left[ x \right]} = \frac{1}{N_r} \sum_{i=1}^{N_r} \left( x_\text{[DEM]} -  x_\text{[NN]} \right)^2 \,\text{,}
\end{align}
where $N_r$ is the number of cases used in the training set, \emph{i.e.}:
\begin{align}
N_r = \text{floor}\left(73,902 \times r / 100\right) \,\text{.}
\end{align}
\begin{figure}[ht]
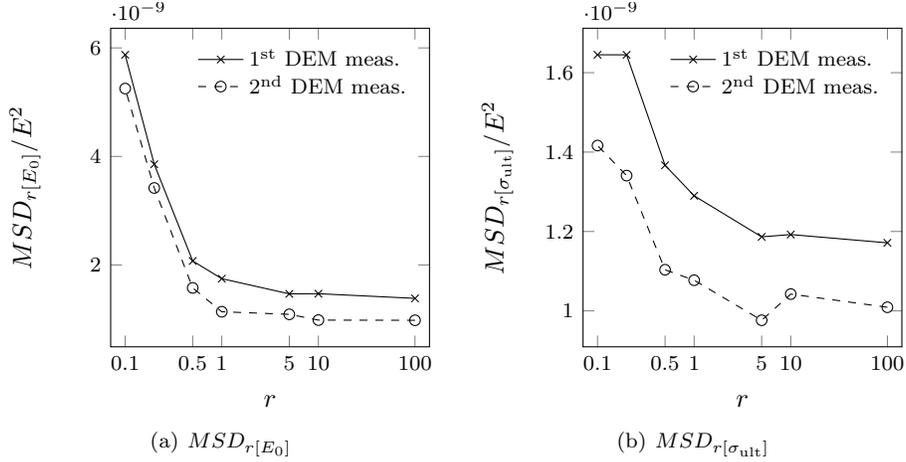

	\centering
	\subfigure[$MSD_{r \left[ E_\text{0} \right]}$]{\includeexternaltikzfigure{figures}{nn_errors_E_percentage}{.}}\hfill
	\subfigure[$MSD_{r \left[ \sigma_\text{ult} \right]}$]{\includeexternaltikzfigure{figures}{nn_errors_q_percentage}{.}}
	\caption{Variation of the mean squared deviations between NN and DEM estimations, respect to the size of training dataset, for the parameters $E_\text{0}$ and $\sigma_\text{ult}$.
	Dashed lines represent the mean squared deviation after the repetition of the most unlikely DEM simulations, whereas solid lines regard the first results.
	The neural networks used for each training dataset $r$ are described (see Table \ref{tab:nns})}
	\label{fig:MSDs}
\end{figure}

Figure~\ref{fig:MSDs} shows how the performance of the NN is barely affected by the size of the training dataset, even when this is drastically reduced.
The networks used in Fig.~\ref{fig:MSDs} are defined in Table~\ref{tab:nns}.
With only $1\%$ of the potential cases (around $700$ DEM experiments), the network NN4 was able to predict the Duncang-Chang model parameters for the test dataset ($18,476$ samples) with almost the same accuracy as NN1.
Thus, we conclude that it is possible to train a NN that accurately predicts the Duncan-Chang parameters from PSDs by just using a dataset with less than one thousand DEM simulations.
It is also important to remark that, to predict the model parameters for a new PSD would take more than 1 hour of computing time, using a DEM model analogous to the ones used in this work, whereas, using an already trained NN the time is in the order of the microseconds.

As it can be seen in Fig.~\ref{fig:MSDs}, the networks' accuracy dropped for networks that were trained with less than 1\% of the potential training samples.
For these small datasets, the accuracy of the NN prediction tended to the value obtained when the sample means of $E_\text{0}$ and $\sigma_{ult}$ are used as estimations.

\subsection{Neural network robustness with respect to noisy DEM training data} \label{sec:noise}

As it can be observed in Fig.~\ref{fig:Discrepancies}, despite the good agreement between NN and DEM predictions for most of the test cases, the discrepancies were relatively high in some of them.
Using network NN1 (see Table~\ref{tab:nns}), the maximum absolute discrepancies were $\Delta_{E_\text{0},\text{max}} = 0.227$ and $\Delta_{\sigma_\text{ult},\text{max}} =0.412$.
Nevertheless, a high discrepancy just means that DEM and NN estimations do not agree, but does not necessarily imply that the NN prediction is wrong.

In order to determine whether these discrepancies were due to inability of the NN to predict the DEM estimation or to unlikely estimations of the model parameters from the DEM, we repeated the virtual triaxial testing in the cases with highest discrepancies.
We considered the networks that were trained with 1\%, 5\%, 10\% and 100\% of potential training cases (networks NN1, NN2, NN3 and NN4 in Table~\ref{tab:nns}).
We identified the $100$ predictions with the highest deviation in $E_\text{0}$ and $\sigma_\text{ult}$, for the networks NN2, NN3 and NN4, and the worst $1,000$ deviations for the network NN1.
Many of them overlapped, so we finally selected around $1,450$ experiments to repeat.
It is worth emphasizing that we did not repeat some of the cases to train the NN again in order to achieve a better matching with different data, the NNs remained unchanged.

\begin{figure}[ht]
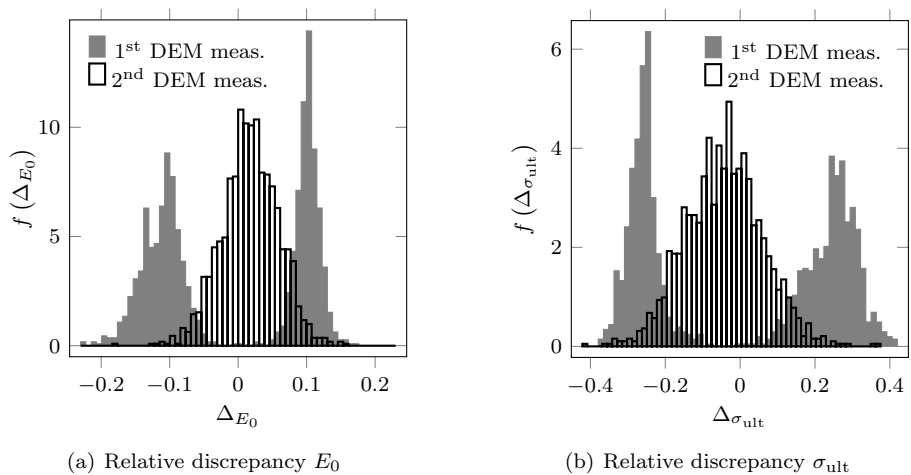

	\centering
	\subfigure[Relative discrepancy $E_0$]{\includeexternaltikzfigure{figures}{histogram_error_E0_after_repetition}{.}}\hfill
	\subfigure[Relative discrepancy $\sigma_\text{ult}$]{\includeexternaltikzfigure{figures}{histogram_error_Qult_after_repetition}{.}}
	\caption{Discrepancies between NN predictions and first and second DEM measurements in the cases that had given the highest discrepancies when comparing the first DEM measurement to the NN predictions. Results obtained with the network NN1 (see Table~\ref{tab:nns})}
	\label{fig:Discrepancies_after_rerpetition}
\end{figure}

After the repetition of these simulations, the relative discrepancies were considerably reduced in most of these cases (see Figs.~\ref{fig:MSDs} and \ref{fig:Discrepancies_after_rerpetition}). 
The standard deviation of the discrepancies over the whole test dataset were also reduced: $s_{\Delta_{E_\text{0}}}$ went from $0.046$ to $0.037$ and $s_{\Delta_{\sigma_\text{ult}}}$ went from $0.115$ to $0.090$.
This proves that for the repeated cases the first DEM measurement was very unlikely, whereas the NN prediction was much more accurate.

This result does not come as a surprise: As already highlighted in some recent works (see, \textit{e.g.}, \cite{rolnick2017deep}), NNs are robust to a certain extent respect to mislabeled or noisy training data.
In the context of this work, this can be understood based on the fact that the NN was trained using  datasets that contain many test cases corresponding
to PSDs that are very close to those being troublesome.
Therefore the NN downplays the contribution of outliers.
Thus, as it was done in this work, the trained NN can also be used as a tool for identifying unlikely estimations of the DEM.

\begin{figure}[ht]
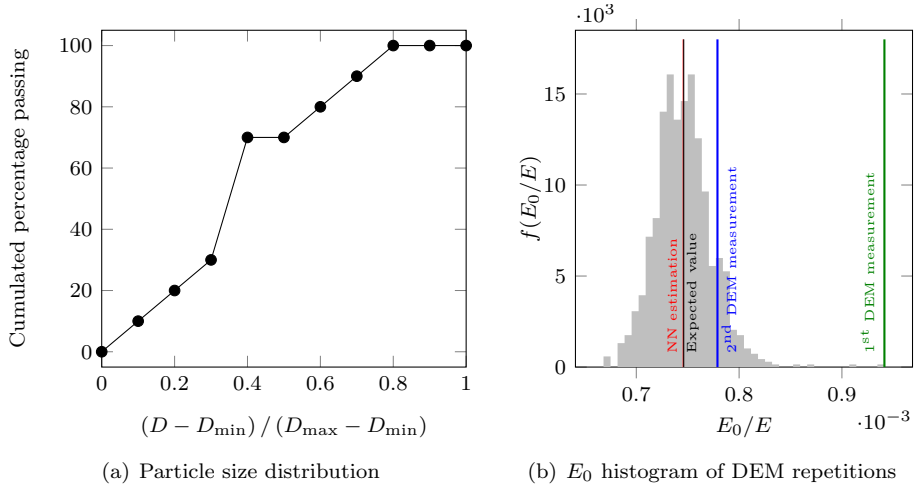

  \centering
	\subfigure[Particle size distribution]{\includeexternaltikzfigure{figures}{PSD_specialcase}{.}\label{fig:PSD59861}}\hfill
	\subfigure[$E_\text{0}$ histogram of DEM repetitions]
	{\includeexternaltikzfigure{figures}{histogram_specialcase}{.}\label{fig:E0histogram59861}}
  \caption{The case 59861 showed one of the highest relative discrepancies between DEM and NN estimations.
	The packing generation and triaxial test were repeated 1000 times for that specific PSD.
	Its PSD and the histogram of predicted $E_\text{0}$ values are shown together with the NN estimation and first and second DEM measurements}
  \label{fig:59861}
\end{figure}

In order to further support this claim, one of the cases with the highest discrepancy between the NN estimation and the DEM measurement (PSD case 59861, see Fig.~\ref{fig:PSD59861}), was more thoroughly analyzed.
This PSD was used to randomly generate $1000$ new packings to be subjected to DEM triaxial compression.
The stress-strain curves were fitted to Duncan-Chang model, generating statistical samples of $E_\text{0}$ and $\sigma_\text{ult}$ values for this single PSD.
With such large samples we could estimate the expected values of the model parameters for PSD 59861 from the samples means.
Focusing on the tangent stiffness $E_\text{0}$, the obtained sample mean was $\bar{E}_\text{0} / E = 7.461 \cdot 10^{-4}$, its standard deviation was $s_{E_\text{0}} / E = 2.791 \cdot 10^{-5}$ ($CoV_{E_\text{0}} = 0.037$) and the minimum and maximum values were $E_\text{0,min} / \bar{E}_\text{0} = 0.895$ and $E_\text{0,max} / \bar{E}_\text{0} = 1.262$, respectively.

Figure~\ref{fig:E0histogram59861} shows the histogram of $E_\text{0}$ for these 1000 triaxial tests.
The value estimated in the first DEM test was $E_\text{0,[DEM]} / E = 9.413 \cdot 10^{-4}$.
Therefore, the first DEM measurement provided very unlikely model parameters ($\lvert E_{\text{0,[DEM}_\text{0}\text{]}} - \bar{E}_\text{0} \rvert = 6.992\, s_{E_\text{0}}$) and this is the reason why the discrepancy with NN estimation was so large.
When the experiment was repeated for a second time, the new DEM estimation was $E_{\text{0,[DEM}_\text{1}\text{]}} = 7.790 \cdot 10^{-4}$, which is considerable closer to the expected value ($\lvert E_{\text{0,[DEM}_\text{1}\text{]}} - \bar{E}_\text{0} \rvert = 1.175\, s_{E_\text{0}}$).
In contrast, the NN (which was trained from noisy data) predicted a tangent stiffness value of $E_\text{0,[NN]} / E = 7.456 \cdot 10^{-4}$, which is really close to the expected value ($\lvert E_{\text{0,[NN]}} - \bar{E}_\text{0}\rvert = 0.019\, s_{E_\text{0}}$).

\section{Conclusions} \label{sec:conclusions}
We selected $92,378$ Particle Size Distributions, PSD, lying within two particle sizes.
We performed virtual triaxial tests with the DEM on samples that followed these PSDs.
We fitted the resulting stress-strain curves to Duncan-Chang hyperbolic models, gathering a statistical sample of the two model parameters, namely, $E_\text{0}$ and $\sigma_{\text{ult}}$.
We found variations of these parameters across the statistical sample that are not easily associated to the PSD.
The parameters followed bell-shaped distributions.
In the case of $E_\text{0}$, $CV_{E_{\text{0}}}=s_{E_{\text{0}}}/\bar{E}_{\text{0}}=0.11$ and  $E_{\text{0},\text{max}} / E_{\text{0},\text{min}} = 2.4$.
In the case of $\sigma_{\text{ult}}$,  $CV_{\sigma_{\text{ult}}}=s_{\sigma_{\text{ult}}}/\bar{\sigma}_{\text{ult}}=0.13$ and $\sigma_{\text{ult},\text{max}} / \sigma_{\text{ult},\text{min}} = 3.4$.\\
Because of the finite number of particles used in each experiment ($20,000$), the parameters obtained from a single DEM simulation may fluctuate to some extent from the expected values (with coefficients of variation for measurements of $CV^M_{E_\text{0}} = s_{E_\text{0}} / \bar{E}_\text{0} \simeq 0.05$ and $CV^M_{\sigma_\text{ult}} = s_{\sigma_\text{ult}} / \bar{\sigma}_\text{ult} \simeq 0.10$).\\
We compared DEM measurements to common statistical and geotechnical descriptors derived from the PSD but did not find any correlation.
More precisely, we used the coefficient of uniformity, the coefficient of curvature, the mean size, the standard deviation, the skewness and the excess kurtosis.
In contrast, by using a Neural Network, NN, trained with a dataset generated through DEM simulations, we were able to predict the expected model parameters for each experiment with high accuracy.
The input for this NN was directly the PSD and the output was the model parameters.
We tried several NN architectures.
$20$\% of the dataset was used to test the ability of networks to anticipate the model parameters.
The size of the training dataset varied between $0.1$\% and $100$\% of the remaining DEM experiments.

We observed that the maximum accuracy is similar to the precision of measurement.
This precision is achieved with a training dataset of $1$\% of the potential training cases (about $700$ DEM simulations).
This means that using NNs trained with less than one thousand triaxial experiments it is possible to predict accurately the macroscopic mechanical behavior of granular materials by just using their PSD.

We also observed that the largest discrepancies between NN predictions and DEM measurements occurred precisely when the DEM experiments led to unlikely values in the first simulation.
Therefore the NN was also useful to identify unlikely DEM results.
The key to achieve more accurate estimations seems to be the reduction of the data noise.

The PSD often affects the mechanical behavior of granular materials.
There must exist relationships linking the mechanical behavior to the PSD that are hidden to the naked eye. Nor even using statistical or geotechnical descriptors that may quantify the PSD to some extent, relationships could be established. In contrast, neural networks were capable of finding those relationships.
This research opens a way to address other problems (\textit{e.g.}, different stress-strain paths or sample preparation procedures), with the objective of better understanding the relationship between PSDs and the macroscopic behavior.
The great advantage of the combination of DEM with NN is that we can know much more by simulating much less. 

\section*{Acknowledgments}
P.\ Antolin was partially supported by the European Research Council through the H2020 ERC Advanced Grant 2015 n.694515 CHANGE,
and by the Swiss National Science Foundation through the project ``Design-through-Analysis (of PDEs): the litmus test'' n.40B2-0\_187094 (BRIDGE Discovery 2019).

\bibliography{mybibfile}

\end{document}